\documentstyle[prc,aps,epsfig]{revtex}
\begin{document}
\draft
\title{Mediating states in $^{180}$Ta}
\author{Petr Alexa}
\address{Department of Physics and Measurements, Institute of
Chemical Technology, \\ 
Technick\' a 5, CZ-166 28 Prague 6, Czech Republic}
\date{\today}
\maketitle

\begin{abstract}
Possible theoretical electromagnetic paths between the ground state 
and the isomeric state at 75.3 keV in $^{180}$Ta are discussed in the 
framework of the two-quasiparticle-plus-phonon model and the standard 
axially-symmetric rotor model including Coriolis mixing.
Experimental transition rates from the isomeric state to the ground state
via observed mediating states are compared to the theoretical ones.
\end{abstract}

\section{Introduction}
$^{180}$Ta is the only nucleus present in nature in
an isomeric state ($9^-$) at an energy of 75.3 keV. Its solar abundance
is very small (2.48, normalized to $10^{12}$ for Si), thus presenting a 
challenge for different nucleosynthesis models (see \cite{Alexa-astro} for a 
detailed discussion of proposed production scenarios). 

In the s-process site thermal photons may excite higher-lying states in 
$^{180}$Ta which then decay back either to the $1^{+}$ ground state or to 
the $9^{-}$ isomer. To find these mediating states (MS) 
Belic {\it et al.} \cite{Bel,Belnew} used the Stuttgart 
Dynamitron facility with both enriched ($5.6\%$) and natural Ta targets. 
Irradiations were performed for bremsstrahlung endpoint energies 
$E_0 = 0.8$--$3.1$ MeV. Depopulation of the isomer was observed down to 
$E_0 \approx 1.01$ MeV. This means that the lowest MS may have an 
excitation energy $E_{MS} = 1.085$ MeV (above the ground state).
The experimental total integrated depopulation
cross section $I_D$ then turns out to be $(5.7\pm 1.2)$ eV fm$^2$.
Higher lying mediating states (below 2 MeV) were found at $E_{MS} =
1.30$ MeV ($I_D = 27$  eV fm$^2$), 1.51 MeV ($I_D = 24$  eV fm$^2$),
1.63 MeV ($I_D = 70$  eV fm$^2$), and 1.93 MeV ($I_D = 111$ eV fm$^2$).
The structure, spin and parity of the MS remain unknown, calculations in
the framework of the two-quasiparticle-plus-phonon model (TQPM) 
\cite{Soloviev1} failed to reproduce 
their energies as well as $I_D$'s (MS found for $E_{MS}>2.4$ MeV).

\section{Model description}
For a theoretical description of $^{180}$Ta we use the standard axially
symmetric rotor model including Coriolis mixing \cite{Kvasil1}. The 
intrinsic degrees of freedom are described 
in the framework of the TQPM \cite{Kvasil2}. 
The model Hamiltonian is given by a
deformed axially symmetric average field (Nilsson potential with
parameters from \cite{Soloviev}), monopole pairing 
interaction (proton and neutron gaps from \cite{Soloviev}) and a 
long-range residual multipole-multipole interaction:
\begin{equation}
\label{Hmm}
\hat H_{\rm mm} = -\frac{1}{2} \sum_{\lambda = 2,3;\mu }
\kappa_0^{(\lambda\mu)} \hat Q_{\lambda \mu }
\hat Q_{\lambda -\mu} \ .
\end{equation}
The strength constants, $\kappa_0^{(\lambda\mu)}$, are
fitted to experimental energies of the $\lambda \mu $--vibrational
states of the even-even core or taken from systematics.

The model Hamiltonian is treated in the BCS approximation.
One-phonon even-even core excitations are obtained using the
standard RPA \cite{Soloviev}. All terms of the two quasiparticle 
interaction in the model Hamiltonian corresponding to the 
neutron-proton multipole-multipole interaction are replaced by a 
diagonal Gaussian force with central, spin-spin, Majorana and 
Majorana spin-spin components with parameters from \cite{Boisson}.

The intrinsic model wave-functions are composed of
one-neutron-quasi\-particle plus one-proton-quasiparticle and  
one-neutron-quasiparticle plus one-proton-quasiparticle 
plus phonon components:
\begin{equation}
\mid \psi_{K} \rangle = \left \{ \sum_{np}
C_{npK} \alpha_n^{\dagger} \alpha_p^{\dagger} +
\sum_{npg} D_{npgK} \alpha_n^{\dagger} \alpha_p^{\dagger} Q_g^{\dagger}
 \right \} \mid \rangle,
\end{equation}
where the parameters $C_{npK}$ and $D_{npgK}$ are determined using the
variational principle. 

\section{Results and conclusions}
In the model, vibrational admixtures in neutron-proton wave functions
can be calculated. Taking into account the Coriolis mixing (without any 
attenuation) enables us to 
predict possible transitions between the ground state and the 
isomer. Recently, we improved our calculations \cite{Alexa}.
The model space comprises now 260 lowest intrinsic 
two-quasiparticle states with one-phonon components and corresponding
rotational bands.
For the calculation of electromagnetic transitions (and
branching ratios of the $^{180}$Ta levels to the ground state and the 
isomer), $e_{\rm p,eff}($E1$) = 1.2 e$ and $e_{\rm n,eff}($E1$) = 0.8 e$,
$e_{\rm p,eff}($E2$) = e$ and $e_{\rm n,eff}($E2$) = 0.2 e$,
$e_{\rm p,eff}($E3$) = e$ and $e_{\rm n,eff}($E3$) = 0.2 e$,
$g_{\rm s,red} = 0.7$ and $g_{\rm R}= 0.26$ were used. Theoretical
energies were replaced by known experimental energies 
\cite{Dracoulis1,Dracoulis2,Wendel} 
and internal conversion was taken into account. 

\begin{figure}
\epsfxsize = \textwidth
\centerline{\epsfbox[66 72 528 270]{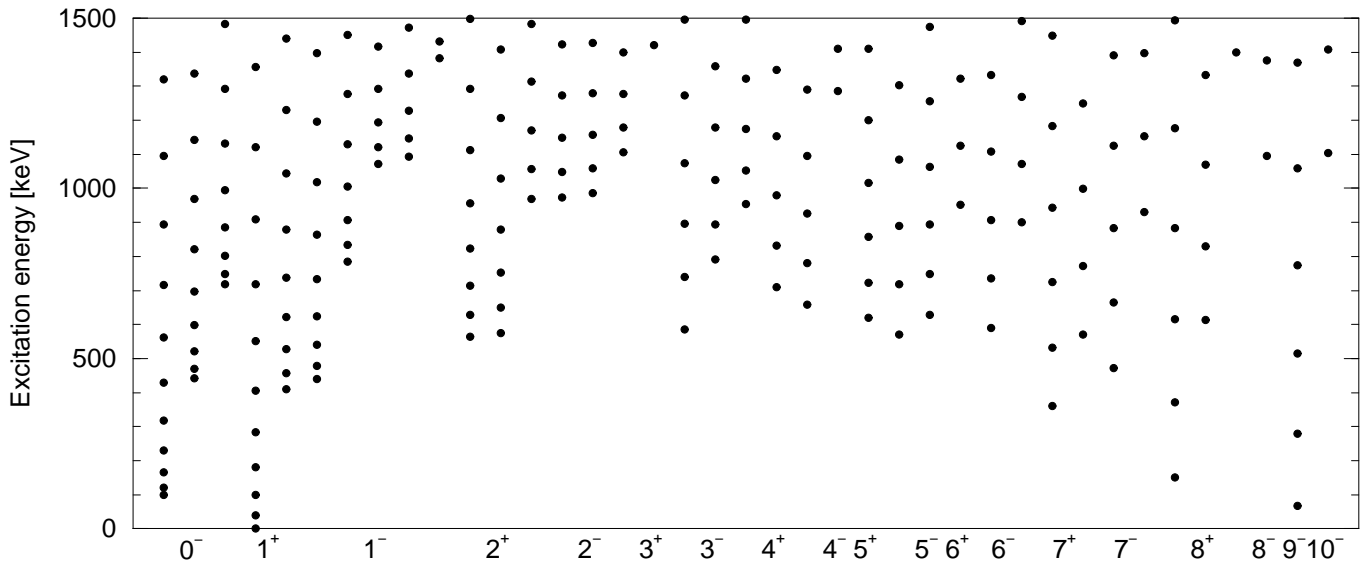}}
\caption{Theoretical energies vs. intrinsic spin projection $K$ and
parity for low-lying states in $^{180}$Ta.}
\label{Tasp}
\end{figure}

\begin{figure}
\epsfxsize = 8.5cm
\centerline{\epsfbox[38 10 502 497]{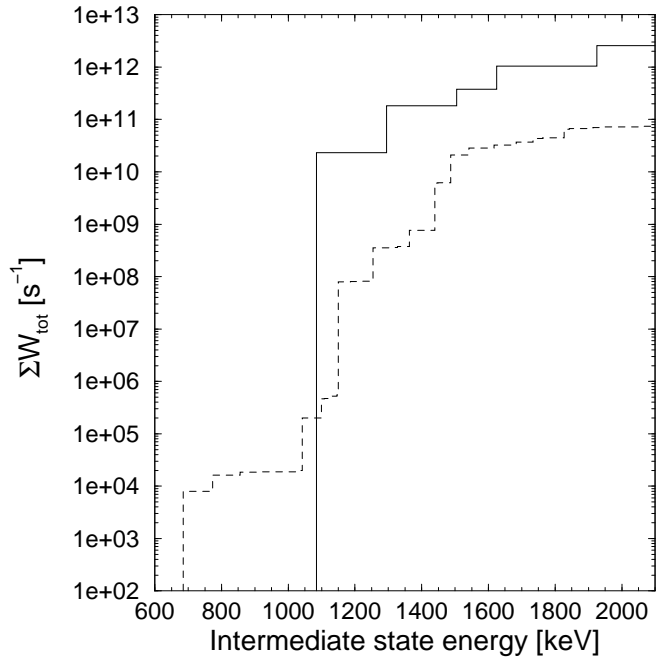}}
\caption{Summed experimental transition rates $\sum W_{tot}$ (solid line) in 
comparison with the results of the TQPM+PRM calculations (dashed line).}
\label{wtot}
\end{figure}

The main results can be summarized as follows:
\begin{enumerate}
\item
There are no MS at low energies (below 600 keV).
This is demonstrated in Fig.~\ref{Tasp} where theoretical
energies vs. intrinsic spin projection $K$ and parity are plotted.

\item
The total experimental transition rate $W_{tot}$ for the process \lq isomer 
$\rightarrow $ MS $\rightarrow $ ground state\rq \ can be calculated from
\begin{equation}
W_{tot} = \frac{I_D}{\hbar} \cdot 
\left(\frac{E_{MS}-75.3\mbox{ keV}}{\pi \hbar c}\right)^2
\end{equation}
and compared with the theoretical transition rate
\begin{equation}
W=\frac{8\pi }{\hbar} \sum_{Xl} \frac{l+1}{l[(2l+1)!!]^2} 
\cdot \left(\frac{E_{MS}-75.3\mbox{ keV}}{\hbar c}\right)^{2l+1} \cdot 
B_{eff}(Xl) \ ,
\end{equation} 
where we sum over the relevant multipolarities E1, E2, and M1. $B_{eff}(Xl)$
is the effective reduced transition probability for the process
\lq isomer $\rightarrow $ MS $\rightarrow $ ground state\rq . 
In Fig.~\ref{wtot} the summed $W_{tot}$ and $W$ are compared.
The main trend is reproduced but we still fail to account for 1 or 2 orders
of magnitude.
\end{enumerate}

In our theoretical calculations using the TQPM and the standard 
axially-symmetric rotor model including Coriolis mixing low-lying mediating 
states were found for the first time. The lowest mediating state lies at 670
keV, but the transition rate from the isomer to the ground state is very small
(see Fig.~\ref{wtot}). A larger increase of the transition rate was found
in the region around 1.2 MeV near the lowest experimentally observed mediating
state.


\begin{thebibliography}{9}
\bibitem{Alexa-astro} M.\ Loewe, P.\ Alexa, J.\ de Boer, and M.\ W\"urkner,
astro-ph/0207123.
\bibitem{Bel}D.\ Belic, C.\ Arlandini, J.\ Besserer, J.\ de Boer, 
J.J.\ Carroll, J.\ Enders, T.\ Hartmann, F.\ K{\"a}ppeler, H.\ Kaiser,
U.\ Kneissl, M.\ Loewe, H.J.\ Maier, H.\ Maser, P.\ Mohr, P.\ von 
Neumann-Cosel, A.\ Nord, H.H.\ Pitz, A.\ Richter, M.\ Schumann, S.\ Volz, and 
A.\ Zilges, Phys. Rev. Lett. {\bf 83}, 5242 (1999).
\bibitem{Belnew}D.\ Belic, C.\ Arlandini, J.\ Besserer, J.\ de Boer, J.J.\
Carroll, J.\ Enders, T.\ Hartmann, F.\ K\"appeler, H.\ Kaiser, U.\ Kneissl,
E.\ Kolbe, K.\ Langanke, M.\ Loewe, H.J.\ Maier, H.\ Maser, P.\ Mohr, P.\
von Neumann-Cosel, A.\ Nord, H.H.\ Pitz, A.\ Richter, M.\ Schumann, F.-K.\
Thielemann, S.\ Volz, and A.\ Zilges, Phys. Rev. C {\bf 65}, 035801 (2002).
\bibitem{Soloviev1} V.G.\ Soloviev, A.V.\ Sushkov, and N.Yu.\ Shirikova,
Phys. At. Nucl. {\bf 64}, 1199 (2001).  
\bibitem{Kvasil1} J.\ Kvasil, Czech. J. Phys. {\bf B31}, 1376 (1981).
\bibitem{Kvasil2} J.\ Kvasil, R.K.\ Sheline, V.O.\ Nesterenko, 
I.\ H\v rivn\' a\v cov\' a, and D.\ Nosek, Z. Phys. {\bf A343},
145 (1992).
\bibitem{Soloviev} V.G.\ Soloviev, Theory of Complex Nuclei,
Pergamon Press, Oxford 1976. 
\bibitem{Boisson} J.P.\ Boisson, R.\ Piepenbring, and W. Ogle,
Phys. Rep. {\bf 26}, 99 (1976).
\bibitem{Alexa} P.\ Alexa, I.\ H\v rivn\'a\v cov\'a, and J.\ Kvasil, Acta
Phys. Pol. {\bf B30}, 1323 (1999).
\bibitem{Dracoulis1} G.D.\ Dracoulis, S.M.\ Mullins, A.P.\ Byrne,
F.G.\ Kondev, T.\ Kib\'edi, S.\ Bayer, G.J.\ Lane, T.R.\ McGoram, and
P.M.\ Davidson, Phys. Rev. C {\bf 58}, 1444 (1998). 
\bibitem{Dracoulis2} G.D.\ Dracoulis, T.\ Kib\'edi, A.P.\ Byrne, R.A.\ Bark, 
and A.M.\ Baxter, Phys. Rev. C {\bf 62}, 037301 (2000).
\bibitem{Wendel} T.\ Wendel, J.\ Gr\"oger, C.\ G\"unther, A.I.\ Levon, 
P.E.\ Garrett, L.\ Genilloud, J.\ Jolie, J.\ Kern, S.\ Mannanal, N.\ Warr, 
F.\ K\"appeler, G.\ Graw, R.\ Hertenberger, M.\ Loewe, and M.\ W\"urkner,
Phys. Rev. C {\bf 65}, 014309 (2002).
\end{thebibliography}
\end{document}